\documentclass[12pt, preprint]{aastex}
\begin{document}
\title{Is the Galactic Cosmic Ray Spectrum Constant in  Time?}
\author{David Eichler\\
        Physics Department, Ben-Gurion University, Be'er-Sheva 84105, Israel\\
        E-mail: \email{eichler.david@gmail.com}}
         \author{Rahul Kumar\\
        Physics Department, Ben-Gurion University, Be'er-Sheva 84105, Israel\\
        E-mail: \email{rahuliitk@gmail.com}}
\author{{Martin Pohl}$^{a,b}$\\
    {$^a$}Institut f\"ur Physik und Astronomie, Universit\"at
Potsdam, 14476 Potsdam-Golm, Germany\\
{$^b$} DESY, 15738 Zeuthen, Germany\\
        E-mail: \email{pohlmadq@gmail.com}}

\begin{abstract}
   The hypothesis is considered that the present Galactic cosmic ray spectrum is at present softer than its time average due to source intermittency. Measurements of muogenic nuclides underground could provide an independent measurement of the time averaged spectrum. Source intermittency could also account for the surprising low anisotropy  reported by the IceCube collaboration. Predictions for Galactic emission of ultrahigh-energy quanta, such as UHE gamma rays and neutrinos, might be higher or lower than previously estimated.
   \end{abstract}

\section{Introduction}
The measured residence spectrum of Galactic cosmic rays (CR) is $N_r(E)dE\propto E^{-2.7}dE $ from 1 GeV to $E_{knee}  = 3$ PeV, and $E^{-3.3}$ from $E_{knee}$ to $E_{ankle}= 4$ EeV. This is considerably steeper than the theoretical expectation for the injection spectrum  of $N_i(E)dE\propto E^{-2.0}dE $ for diffusive shock acceleration (DSA) by strong shocks. Much of the difference is attributed to energy dependent escape  rate R(E) from the Galaxy $R(E)\propto E^{d}$. However, it is questionable whether the entire difference is due to this energy dependence.  For one thing, the secondary spectrum is about $N_s(E)dE\propto E^{-3.0}dE $ suggesting that $d\sim {0.3}$ (Trotta et al., 2011), and, by implication, $N_i(E)dE\propto E^{-2.4}dE $. Moreover, if $d$ were much more than 0.3 over many decades of energy, then the anisotropy at the highest energy of Galactic CR near 1 EeV, would be much higher than the reported upper bound of of $\sim 10^{-2}$.

There could be many reasons for the injection index to be much steeper than 2.0, but it would in any case be good to know its exact value, as the total energy requirements for Galactic CR are a sensitive  function of it. Moreover, the UHE emission that could be expected from dense environments that offer a thick target to such CR are an extremely sensitive function of the injection index. In this letter we consider the unconventional possibility that the time average residence spectrum is actually flatter than -2.7 by some small amount $\epsilon$. We would then conclude that the  injection index is -(2.4- $2\epsilon$).

The possibility and consequences of time variability of CR have received increased attention (Melott and Thomas, 2011, Erlykhin and Wolfendale, 2010 and references within) following the suggestions that CR flux correlates with cloud cover (Svensmark, Bondo and Svensmark, 1997) and that very high energy CR trigger lightning (Gurevich, Milikh, and Roussel-Dupre 1992). Erlykhin and Wolfendale have specifically calculated the variation in CR intensity due to discrete, local supernovae at various energies. As far as we know, there is circumstantial evidence that cloud cover, lightning, and cosmic ray flux all anti-correlate with solar activity. There is not yet a firm causal link established between cosmic ray flux and lightning or cloud cover.  The possibility remains that they could all be tied to a fourth correlate, however, it would still be a mystery what this fourth correlate would be. In any case, variations due to terrestrial and heliospheric conditions are separate from variations in the local Galactic flux. This paper is concerned with the latter.

 At energies much higher than 1 GeV,  CR production in the Galaxy may be intermittent. Intermittency can lower the anisotropy relative to steady state  during the lulls between the intermittent production episodes (Pohl and Eichler, 2011, Pohl and Eichler, 2012, Kumar and Eichler, 2012).  The recently reported anisotropies and/or upper bounds may be  at least partly  understandable in these terms. This explanation for the unexpectedly low anisotropy  would then predict a slightly steeper present day spectrum than the time averaged spectrum.   

Abundances of terrestrial cosmogenic nuclides (TCN) can set limits on recent time averaged variations of the cosmic ray flux at Earth, though not on short term variability.
They demonstrate that the cosmic ray number flux in the Solar System could not have varied on average by more than about 20 or 30 percent of its presently measured value. (Some  weak constraint on even short term cosmic ray variability is placed by the continued existence of life on Earth for hundreds of millions of years. The possibility remains that astrophysically brief fluctuations had interesting terrestrial impact that fell short of total lethality.)
  
The discreteness of supernovae, assuming they are the main source of CR, should in
any case provide an {\it a priori} case for some time variability in the CR flux at Earth.
Supernovae are distributed in the plane of the Galaxy, and most local cosmic rays come from
sources at a distance commensurate with the halo size, $H$, conventionally taken to be a few kpc.
Beyond $H$, cosmic rays preferentially diffuse out of
the face of the Galaxy rather than across the disk, and their contribution at large distances,
$r$, is exponentially small in $r$ (Pohl and Eichler, 2011).

Fluctuations around the time average should increase as the escape time from the Galaxy
decreases. If the  escape time decreases with energy, $E$, so then does $N(E)$, and the level of fluctuation increases with $E$. This amounts to fluctuations in the spectral index.
Taking for example the rate of supernovae, $P$, the CR source strength, $Q$, and
diffusion with diffusion coefficient $D$, the average CR flux in the Galaxy (of effective
radius $R$) is
\begin{equation}
N_{\rm av}(E) \simeq \frac{Q\,P\,\tau_{\rm esc}}{\pi\,R^2\,H}
\end{equation}
where $\tau_{esc}=H^2/2D$ is the escape time. A single source provides CRs that are, let us assume for simplicity,  are distributed over a sphere of radius $r(t) = \sqrt{D\,t}$, where $t$ is the time since
CR release. For $r(t)\le H$, the density of these new CRs therefore is
\begin{equation}
N_{\rm s}(E) \simeq \frac{3\,Q}{4\pi r(t)^3}=\frac{3\,Q}{4\pi\,(D\,t)^{3/2}}\ .
\end{equation}
If the supernova is at distance $r$, one expects a  spike in CR flux of relative amplitude
\begin{equation}
\eta=\frac{N_s}{N_{\rm av}} \simeq \frac{3R^2\,H}{4r^3\,P\,\tau_{\rm esc}}\ .
\end{equation}
For GeV-band CRs $P\,\tau_{\rm esc}\approx 10^6$, and therefore we
have to be within
\begin{equation}
r_c(\eta)\simeq (100\ {\rm pc})\,\eta^{-1/3}\,\left(\frac{D(E)}{D({1\ \rm GeV})}\right)^{1/3}
\end{equation}
of the source to expect a spike of amplitude $\eta$. 
Note that $r_c$ is very weakly dependent
on CR energy. The duration of a large fluctuation ($\eta \gtrsim 1$) CR event is about
\begin{equation}
T\simeq \frac{r_c^2}{D(E)} \simeq 2\,{\tau_{\rm esc}}\,\frac{r_c^2}{H^2}
\simeq (6\cdot 10^4\ {\rm yr})\,\left(\frac{\eta\,D(E)}{D({1\ \rm GeV})}\right)^{-1/3}\ .
\end{equation}
The frequency of such CR events is
\begin{equation}
f\simeq P\,\frac{r_c^2}{R^2}\simeq 10^{-4}\,P\,
\left(\frac{D(E)}{\eta\,D({1\ \rm GeV})}\right)^{2/3}
\end{equation}
and the fraction of time, $\eta$, that one can expect to experience an event is
given by the product of frequency and duration,
\begin{equation}
\eta =f\,T\simeq 0.06\,\eta^{-1}\,(P\cdot 100\,{\rm yr})
\,\left(\frac{D(E)}{D({1\ \rm GeV})}\right)^{1/3}
\end{equation}
Thus, significant events with $\eta > 1$ are somewhat exceptional, but not unexpected. They are likely to be accompanied by
spectral distortions because, at a fixed distance from the source,  high-energy CRs arrive earlier than
low-energy CRs.


    In this letter, we consider the possibility that the average residence index in the Galaxy is -(2.7 - $\epsilon$), $\epsilon \sim 10^{-1}$, that the escape rate exponent $\delta$ is in fact 0.3 + $\epsilon$, and  that the time average CR injection  index is -(2.4-$2\epsilon$) rather than -2.4. This time averaged spectrum is slightly "propped up" by nearby supernovae, which introduce freshly accelerated populations.  That the present day spectrum  is then as steep as  $E^{-2.7}$, according to the present hypothesis,  due to the paucity of recent, nearby supernovae.

  \section{Cosmogenic Nuclei Diagnostics}
   We suggest that the variation of the CR spectral index over time in the GeV to TeV range could affect, at detectable levels, the rate of muogenic nuclei  production in deep rock formations that exceed 3 Myr in age.
   Studying nuclear cosmogenesis at and below the Earth's surface has the advantage, in  searching for evidence of hard transient sources, that the overlying material filters out much more of the contribution from low energy primaries than from the high energy ones.  In particular, muons typically arrive from CR primaries of  $0.03\le E\sim 1$  TeV, and would be more prevalent, relative to the total cosmic ray flux, for harder spectra.
   
    The depth would need to be large enough to be unaffected by possible uncertainties in the erosion rate at the surface.  Variations in the CR intensity (due to nearby supernovae, or due to long term magnetic field change in the sun or Earth) within the past 3 Myr could possibly be tested by $^{10}$Be abundance, to measure the exposure age - and, at the same geographical location, at depths of at least several meters, to isolate the muogenic component. Alternatively, measurements deep enough to give reliable exposure ages, independent of uncertainties in erosion rates, would give an absolute measurement of the past muon flux. The cross section for muogenic $^{53}$Mn  (half life 
   3.7 Myr,  exponential decay time 5.3 Myr.) from an iron target is 3.8 mb at $E_{\mu}=190$ GeV. (Heisinger et al, 2001).  The production rate per target atom, with a muon flux that is the same as at sea level, is then $10^{-21}$/yr (equation 17 of Heisinger, et. al, 2001). Here we have allowed for the fact that the mean energy at sea level is only $\sim 10$ GeV, and that the cross section for muogenic $^{53}$Mn production varies as $\sim E_{\mu}^{0.86}$. The production flux remains within an order of magnitude of the sea-level value down to depths of $10^4$g/cm$^2$, and analysis of the $^{53}$Mn content of subterranean iron ore would give a straightforward measurement of the exponentially-weighted average muon flux over the past 5 Myr. Comparison with muogenic $^{10}$Be (half life 1.6 Myr, production cross section on O at 190 GeV of 0.094 mb), could help date the contribution of recent nearby supernovae within the residence time in the disk ($\sim 3 $ Myr). Evidence for a recent nearby supernova, occurring about 2.8 Myr ago, has been presented by Knie et al.(2004).

   Measuring changes in the CR spectrum that are restricted to energy range $E_p\gg 1$ TeV would be  more difficult, so a time averaged CR flux that is well above the present only at $E_p\ge $1 TeV (such as might result from CR injection by ultrarelativistic shocks with Lorentz factors exceeding 100) would be hard to rule out. Muons from primaries at $E_p\gg$ 1 TeV dominate the total flux only at depths of $\gtrsim 10^5$ g cm$^{-2}$, where the nuclear muogenesis rate per target atom is only about $10^{-23}$/yr. A net collection of $10^6$ $^{53}$Mn atoms, accumulating over $10^{6.5}$ yr,  would thus require a purified sample of 10$^{22.5}$  iron nuclei, i.e. several grams.

  The dependence of muon secondary flux, and its depth dependence, can be estimated with a simple analytic model:
   Muons at sea level come mainly from primaries at $E\gtrsim 100 GeV$, which create center-of-mass Lorentz factors that are larger than  the ratio of the muon flight time to its rest frame lifetime. A detailed account of the physics is given in Dorman (2004), and a look-up table for muon production as a function of primary energy is given by Atri and Melott (2011). ; 
    for the following discussion,  a rough analytic estimate will suffice, and we take the atmosphere to be isothermal and planar: As primaries create the first generation muons at about 1 interaction depth which we take to be 85 g/cm$^2$ in air,  or about $1/12$ of the total vertical grammage,  1030 g/cm$^2$,  of the atmosphere, this typically occurs at an altitude of $h_* \sim 18/cos \theta$ km, where $\theta$ is the angle of incidence of the primary relative to the zenith.
    From the muon production altitude $h_*$ to altitude h, the proper flight time in the muon rest frame is given by

    \begin{equation}
    \tau(\theta,h)= \int_{h_*}^h d\tau = \int_{h_*}^{h} -dh'/\gamma(h',\theta) c \cos\theta
    \label{dtau}
    \end{equation}
     where 
     \begin{equation}
    \gamma(h,\theta)=\gamma(h_*)-20G(h,\theta)/G(0,0)= \gamma(h_*)-20\left[\int_{h}^{h_*}\rho(h') dh'/\cos \theta \right]/G(0,0)
     \end{equation}
     The above expression assumes that muons lose 20 times their rest energy, $1.05 \cdot 10^8$ eV, per vertical atmospheric grammage $G(0,0)$ - i.e. about 2 MeV per gram over $1.03 \cdot 10^3$
     g cm$^{-2}$. Assuming a density distribution of $\rho(h)=\rho_o exp[-h/h_s]$, neglecting terms of order $exp[-h_*/h_s]$, and integrating equation (\ref{dtau}) from $h_*$ to altitude h, $h\ll h_*$,  we obtain

     \begin{equation}
     \tau(\theta,h) = \frac{h_s/c}{\gamma_*\cos \theta}ln\frac{\left[e^{h_*/h_s}\gamma_*\cos \theta -20\right]}
     {\left[e^{h/h_s}\gamma_*\cos \theta -20\right]}
   \end{equation}
   where  $\gamma_*\equiv  \gamma(h_*)$.  
   For $h_s\sim 8 $ km, a reasonable value for the Earth's atmosphere over the lowest 3 scale heights,  we can write $h_s/c\sim 12 \tau_{\mu}$, where $\tau_{\mu}$  is the muon lifetime.
       The decay factor is then
    \begin{equation}
    e^{-\tau(h, \theta)/\tau_{\mu}}= \left[((e^{h/h_s}-20/\gamma_*\cos \theta)\over(e^{h_*/h_s}-20/\gamma_*\cos \theta)\right]^{12/\gamma_*\cos \theta}.
    \end{equation}

   The muon flux at sea level, $F_{\mu}(\gamma(0),\theta,0)$, can be expressed as
     \begin{equation}
    {d^2 F_{\mu}(\gamma,\mu,0)\over d\gamma d\mu}d\gamma d\mu= {d^2F_{\mu}(\gamma_*,\mu, h_*)\over d\gamma_* d\mu} e^{-\tau(\mu,0)/\tau_{\mu}}d\gamma_*d\mu
   \end{equation}

       Integrating over $\mu$ and over $ \gamma_*$, over the range where $\gamma_*\gtrsim 12/\mu$,  and assuming that the muons have a spectrum of $ {d^2F_{\mu}(\gamma_*,\mu, h_*)\over d\gamma d\mu} \simeq  {d^2F_{\mu}(1,\mu, h_*)\over d\gamma d\mu}\gamma_*^{-2-p}\equiv C\gamma_*^{-2-p} $, we obtain for the all sky muon flux at sea level, $F_{\mu}$,

   \begin{equation}
      F_{\mu}(0)\simeq C\int_0^1 d\cos \theta \int_{20/\cos\theta}^{\infty}\gamma_*^{-2-p} \left[(1-20/\gamma_* \cos \theta)\over(e^{h/h_s}-20/\gamma_* \cos\theta)\right]^{12/\gamma_* \cos \theta}d\gamma_*
   \end{equation}
   where the lower limit of the integral is determined by the condition that the integrand be positive.

   When $\gamma_*\gg h_*/c \tau_{\mu}\cos\theta \sim 27 /\cos\theta$, $ \tau(h, \theta)/\tau_{\mu}\ll 1$, and the decay factor is about unity. 
     Making the approximation that the decay factor is close to 0 when $\gamma_*\le h_*/h_s \sim 27/\cos\theta$, and 1 when $\gamma_*\ge 27 /\cos\theta$, we can estimate the above integral as
   \begin{equation}
      F_{\mu}(0)\simeq C\int_0^1 d\cos \theta \int_{27/\cos \theta}^{\infty}\gamma^{-2-p} d\gamma_*
   =  [1/(p+1)(p+2)(27)^{p+1}]C
   \end{equation}

   Thus,  decreasing p from 0.7 to 0.6  
   with ${dF_{\mu}\over d\gamma_*}(1 GeV )$ remaining constant, increases the flux near sea level, $g\equiv G(h,0)/G(0,0)\sim 1$, by a factor of $(27)^{0.1}(1.7/1.6)(2.7/2.6)\sim 1.53$ .  
   This is a good enough estimate for considering the general feasibility of using deep muogenic isotopes to measure the past cosmic ray spectrum.

   Deep underground, where
    $g\gg 27/20$, the incident muons need to begin with a minimum $\gamma_*$ of  $20g/\cos \theta$ in order to survive energy loss, even though their decay is negligible.  The flux can then be approximated as
     \begin{equation}
      F_{\mu}(g)\simeq \int_0^1 d\cos \theta \int_{20g/\cos \theta}^{\infty}\gamma_*^{-2-p} d\gamma_*.
   \end{equation}

   A change in the spectral index -2-p from -2.7 to -2.6 then lead to a change in the all sky muon flux by a factor of $(20g)^{0.1}(1.7/1.6)(2.7/2.6)\sim 1.5g^{0.1}$.

\section{Predicted Variation in Spectral Index}

\begin{figure}
\begin{center}
\includegraphics[scale=0.5]{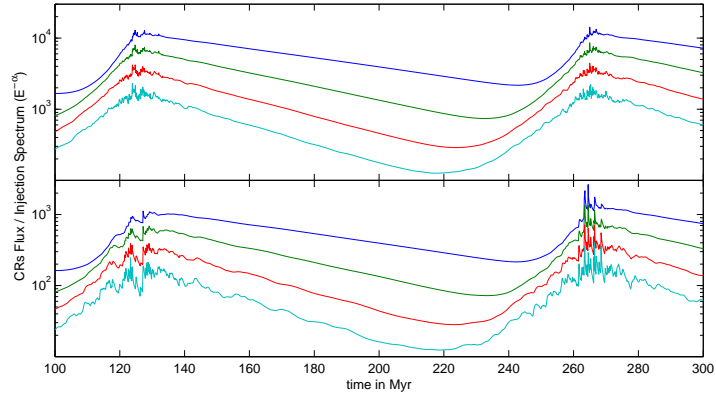}
\end{center}
\label{flux}
\caption{CRs flux variation at four different energies for isotropic diffusion in disk geometry, as sun crosses spiral arms at about 120 Myr (relative time) and 260 Myr. Diffusion coefficient for CRs of energy E is taken to be $6 \times 10^{27} \left(E/ 1 GeV\right)^{1/3} cm^2 s^{-1}$ and disk boundary is at 1 Kpc. Top  and bottom panels are for source rates $10^{-2} $ and $ 10^{-3}$ yr$^{-1}$  respectively. The blue, green, red, and cyan lines are for 1 GeV, 10 GeV, 100 GeV, and 1 TeV respectively, in each panel.}
\end{figure}

\begin{figure}
\begin{center}
\includegraphics[scale=0.5]{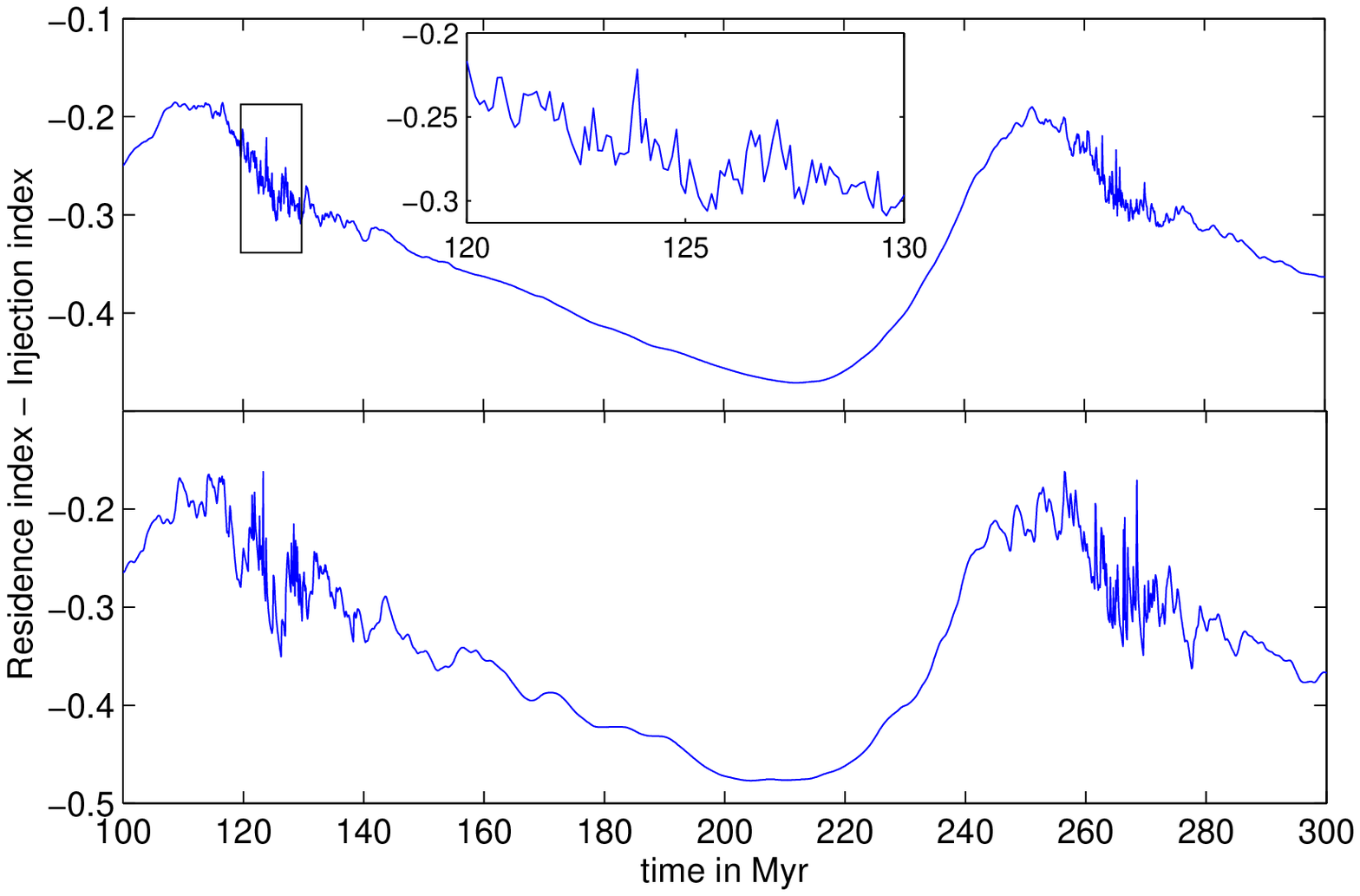}
\label{variation_d}
\caption{Temporal variation of difference in residence and injection indices, considering the CRs spectrum between 1 GeV and 1 TeV (figure \ref{flux}). As in the figure \ref{flux}, top and bottom panels are for source rate $10^{-2}$ and $10^{-3}$ yr$^{-1}$ respectively. Sub-panel in the top panel shows the variation between 120 Myr and 130 Myr (a period when Earth lies just inside a spiral arm) on a smaller time scale (0.1 Myr).}
\end{center}
\end{figure}

We now consider how much the observed spectral index is likely to vary with time at Earth.  We have calculated an expected intensity vs. time plot for randomly placed CR sources in the Galaxy, weighted according to location in proportion to local starlight production. The two figures plot the "effective" escape index (EEI) - i.e. the log of the ratio, as a function of energy, of the CR flux to the source spectrum at the same energy - assuming a value for d of 1/3 and assuming that the cosmic ray sources are dominated by bright supernovae.    
This index differential  can vary from ${-0.2}$, within the spiral arm, to ${-0.45}$ between spiral arms.   This is discussed in more detail elsewhere (Kumar and Eichler, in preparation). Over timescales of 3 Myr, the variation could be of the order of 0.05, which would produce a $^{53}$Mn anomaly of $\gtrsim 10$ percent.  This variation would be detectable if the muon flux averaged over the $^{53}$Mn lifetime could be measured to $\sim 10$ percent accuracy at $g\sim 100$.

The solar system now lies between two major spiral arms but inside a minor spiral arm. There are certainly several nearby, recent supernovae, and they and others may have contributed to some fluctuation of the spectral index over the past several million years.  In addition, however, the excursion into and out of major spiral arms, as illustrated in the figures, may introduce variations of as much as 0.1 or larger in the effective escape index around its average value. The timescale of this variation is longer than several million years, and would not be manifest in $^{53}$Mn anomalies, but perhaps would be in longer lived isotopes. If the CR residence time in the halo is larger than the residence time of the Solar System inside a spiral arm, then the time-averaged value of the EEI is perhaps closer to -0.4 than to 0.3, and the implied source spectral index is flatter by 0.2 than implied by an EEI of -0.3. That is, the inferred source spectrum in closer to -2.2 than  -2.4.

\section{Further Discussion}
The expected emission by the Galaxy of ultrahigh-energy neutrinos, typically generated by primaries in excess of $\sim 10^{14}$ eV, would be raised by about a factor of 3 for each change of 0.1 in the EEI, and the contrast between the spiral arms and the spaces between them might be enhanced by an even larger factor. According to figure 1, this factor could be as high as an order of magnitude or more at such high energies. Similar considerations apply to UHE gamma ray emission from young sources.

  The hypothesis that we are living in an unusual era, when the CR flux at high energies is well below its average value in the Galaxy, might be motivated by anthropic considerations: e.g., if a normal CR flux were somehow detrimental to intelligent life.  The present level of ionizing radiation from CR flux is less than the component due to terrestrial radon, so it is hard to see how changes of less than a factor of 2 in the CR flux could have serious astrobiological consequences, though it cannot be ruled out beyond reasonable doubt. At very high energy ($\sim 10^2$ PeV), intermittency can cause much larger fluctuations in the CR intensity, and if high energy airshowers have astrobiological connections (such as lightning, rainfall and nitrate formation at low altitudes) then it is conceivable that the not unexpected  changes in the spectral index of CR   could conceivably affect terrestrial affairs and the development of life and civilization. Even if very high energy CR are produced by the same events as lower en energy CR, changes in the spectral index of order 0.1 are not unexpected, and even this modest change could influence the flux at 100 PeV by nearly an order of magnitude. A recent nearby supernova, for example, could have briefly raised the CR flux enough to have encouraged life (e.g. by encouraging lightning and, in turn amino acid and/or  nitrate formation), or destroyed it by climatic disturbance or excessive radiation and biological mutation rate.

  Neutrino emission from the spiral arms, and UHE gamma ray emission from supernovae remnants are sensitive functions of their respective local CR spectral index. For a given gamma ray flux at $10^2$ MeV, they could set useful limits on the spatial variation of the CR spectral index.

\acknowledgements
We thank Nir Shaviv and Michael Paul for helpful conversations.
MP acknowledges support by the Helmholtz Alliance for Astroparticle Physics, HAP,
funded by the Initiative and Networking Fund of the Helmholtz Association.
DE  and RK acknowledge support from the Israel-U.S. Binational Science Foundation, the Israeli Science Foundation, and the Joan and Robert Arnow Chair of Theoretical Astrophysics.

\section{References}

\noindent Abbasi, R., Abdou, Y., Abu-Zayyad, T., et al. 2012, ApJ, 746, 33

\noindent Abbasi, R., Abdou, Y., Abu-Zayyad, T., et al. 2010, ApJ, 718, L194

\noindent Atri, D. and A.L. Melott,  Radiation Physics and Chemistry, 2011, 80(6), 701

\noindent Blandford, R.D. \& Eichler, D.E., 1987, Physics Reports, 154, 1

\noindent Boley, F.I., 1964, Rev. Mod. Phys.,   , 792

\noindent Dorman, L., "Cosmic rays in the Earth's Atmosphere and Underground", (Springer, 2004)

\noindent Eichler,  D. 1985, ApJ. 294, 40

\noindent Ellison, D.C. and Eichler, D., 1985, Phys. Rev. Lett., 55, 273

\noindent Erlykin, A.D., and Wolfendale, A.W., 2010,
Surveys in Geophysics, Volume 31, Issue 4, 383

\noindent Gosse, J. C. \& Phillips, F. M. 2001, Quaternary Science Reviews,  20,  147

\noindent Gurevich, A.V., Milikh, G.M., Roussel-Dupre, R., 1992, Phys. Lett. A 165, 463.

\noindent Heisinger, B.,  Lal,D., Jull, A.J.T.,  Kubik, P. Ivy-,  Neumaier, S.,
 Knie, K.,   Lazarev, V.,  Nolte, E.  2002, Earth and Planetary Science Letters 200, 345
 
\noindent Kumar, R., \& Eichler, D. 2012, ApJ, submitted

\noindent Levinson, A. \& Eichler, D. 1993, ApJ 418, L386

\noindent Melott, A., Thomas, B.C. 2011, Astrobiology, 11, 343

\noindent Knie, K., Korschinek, G., Faestermann, T., Dorfi, E. A., Rugel, G., Wallner, A., 2004, Phys. Rev. Lett. 93, 171103

\noindent Pohl, M., \& Eichler, D. 2011, ApJ, 742, 114

\noindent Pohl, M., \& Eichler, D. 2012, ApJ, submitted

\noindent Pohl, M., Perrot, C., Grenier, I., Digel, S. 2003, A\& A 409, 581

\noindent Pohl, M.,
\& Esposito, J.~A.\ 1998, ApJ, 507, 327

\noindent Svensmark, H., Bondo T., \& Svensmarks, J. 2009, Geophysical Research Letters, , 2009, 36, L15101,

\noindent Trotta, R.,
J{\'o}hannesson, G., Moskalenko, I.~V., et al.\ 2011, \apj, 729, 106


\end{document}